\documentclass[reprint,amsmath,amssymb,aps,prc,twocolumn,tightenlines,superscriptaddress]{revtex4-1}
\usepackage{graphicx}
\usepackage{units}
\usepackage{epsfig}
\usepackage{bm}
\usepackage{amsfonts}
\usepackage{amsmath}
\usepackage{amssymb}
\usepackage{makecell}
\usepackage[table]{xcolor}
\usepackage{gensymb}
\usepackage[utf8]{inputenc}

\newcommand{\pic}[2][1.0]{\includegraphics[width=#1\columnwidth]{#2}}

\newcommand{\nuc}[2]{\hbox{$^{#1}$#2}}

\begin{document}

\preprint{V. 0.5}

\title{Spectroscopy and lifetime measurements near the proton drip line: $^{26,27,28}$P}


\author{B.~Longfellow}
\affiliation{National Superconducting Cyclotron Laboratory, Michigan State University, East Lansing, Michigan 48824, USA}
\affiliation{Department of Physics and Astronomy, Michigan State University, East Lansing, Michigan 48824, USA}
%
\author{A.~Gade}
\affiliation{National Superconducting Cyclotron Laboratory, Michigan State University, East Lansing, Michigan 48824, USA}
\affiliation{Department of Physics and Astronomy, Michigan State University, East Lansing, Michigan 48824, USA}
\affiliation{The JINA Center for the Evolution of the Elements, Michigan State University, East Lansing, Michigan 48824, USA}

\author{B.~A.~Brown}
\affiliation{National Superconducting Cyclotron Laboratory, Michigan State University, East Lansing, Michigan 48824, USA}
\affiliation{Department of Physics and Astronomy, Michigan State University, East Lansing, Michigan 48824, USA}

\author{D.~Bazin}
\affiliation{National Superconducting Cyclotron Laboratory, Michigan State University, East Lansing, Michigan 48824, USA}
\affiliation{Department of Physics and Astronomy, Michigan State University, East Lansing, Michigan 48824, USA}
\author{P.~C.~Bender}
\altaffiliation{Present address: Department of Physics, University of Massachusetts Lowell, Lowell, Massachusetts 01854, USA.}
\affiliation{National Superconducting Cyclotron Laboratory, Michigan State University, East Lansing, Michigan 48824, USA}
\author{M.~Bowry}
\altaffiliation{Present address: University of the West of Scotland, Paisley PA1 2BE, UK.}
\affiliation{National Superconducting Cyclotron Laboratory, Michigan State University, East Lansing, Michigan 48824, USA}

\author{P.~D.~Cottle}
\affiliation{Department of Physics, Florida State University, Tallahassee, Florida 32306, USA}

\author{B.~Elman}
\affiliation{National Superconducting Cyclotron Laboratory, Michigan State University, East Lansing, Michigan 48824, USA}
\affiliation{Department of Physics and Astronomy, Michigan State University, East Lansing, Michigan 48824, USA}

\author{E.~Lunderberg}
\affiliation{National Superconducting Cyclotron Laboratory, Michigan State University, East Lansing, Michigan 48824, USA}
\affiliation{Department of Physics and Astronomy, Michigan State University, East Lansing, Michigan 48824, USA}

\author{A.~Magilligan}
\affiliation{National Superconducting Cyclotron Laboratory, Michigan State University, East Lansing, Michigan 48824, USA}
\affiliation{Department of Physics and Astronomy, Michigan State University, East Lansing, Michigan 48824, USA}

\author{M.~Spieker}
\affiliation{National Superconducting Cyclotron Laboratory, Michigan State University, East Lansing, Michigan 48824, USA}

\author{D.~Weisshaar}
\affiliation{National Superconducting Cyclotron Laboratory, Michigan State University, East Lansing, Michigan 48824, USA}
\author{S.~J.~Williams}
\altaffiliation{Present address: Diamond Light Source, Harwell Science and Innovation Campus, Didcot, Oxfordshire, OX11 0DE, UK.}
\affiliation{National Superconducting Cyclotron Laboratory, Michigan State University, East Lansing, Michigan 48824, USA}
%

\date{\today}

\begin{abstract}
We report on the observation of excited states in the neutron-deficient phosphorus isotopes \nuc{26,27,28}P via in-beam $\gamma$-ray spectroscopy with both high-efficiency and high-resolution detector arrays. In \nuc{26}P, a previously-unobserved level has been identified at 244(3)~keV, two new measurements of the astrophysically-important $3/2^+$ resonance in \nuc{27}P have been performed, $\gamma$ decays have been assigned to the proton-unbound levels at 2216~keV and 2483~keV in \nuc{28}P, and the $\gamma$-ray lineshape method has been used to make the first determination of the lifetimes of the two lowest-lying excited states in \nuc{28}P. The expected Thomas-Ehrman shifts were calculated and applied to levels in the mirror nuclei. The resulting level energies from this procedure were then compared with the energies of known states in \nuc{26,27,28}P.

\end{abstract}


\maketitle

\section{Introduction}
The properties of the neutron-deficient phosphorus isotopes have prompted a number of experimental and theoretical studies with a focus on both nuclear structure and nuclear astrophysics. \nuc{26}P lies very close to the proton drip line with estimates for its proton separation energy of 140(200)~keV from mass systematics \cite{WANG201236}, 0(90)~keV from the Coulomb energy difference between the ground state of \nuc{26}P and its isobaric analog state in \nuc{26}Si \cite{THOMAS200421}, $-119(16)$~keV from the improved Kelson-Garvey mass relation \cite{TIAN201387}, and 85(30)~keV from parameters of a fit to energies of $1s_{1/2}$ states in lighter nuclei \cite{FORTUNE201796}. The observed narrow momentum distribution and enhanced cross section for one-proton knockout \cite{NAVIN199881} along with significant mirror asymmetry in $\beta$ decay \cite{PEREZ201693} have provided evidence for \nuc{26}P having a proton halo \cite{BROWN1996381,REN199653,GUPTA200228,LIANG200926}. In recent experiments, the first excited state in \nuc{26}P was reported by Nishimura \textit{et al.}~at an energy of 164.4(1)~keV with a 120(9)~ns half-life \cite{NISHIMURA201466} and later confirmed by P{\'e}rez-Loureiro \textit{et al.}~at 164.4(3(2))~keV with a 104(14)~ns half-life \cite{PEREZ201796} with no other excited states observed.

Nuclear structure plays a critical role in the \textit{rp} process due to the low $Q$ values of proton-capture reactions near the proton drip line. The structure of the neighboring phosphorus isotope, \nuc{27}P, is important for the \nuc{26}Si(\textit{p},$\gamma$)\nuc{27}P reaction rate, which has been investigated for its potential role in the nucleosynthesis of the astrophysical $\gamma$-ray emitter \nuc{26}Al during explosive hydrogen burning \cite{WIESCHER1986160,HERNDL199552,GUO200673,TIMOFEYUK200878,JUNG201285,FORTUNE201592,CAGGIANO200164,TOGANO201184,MARGANIEC201693}. Under typical nova conditions, the low-lying $3/2^+$ resonance provides the dominant contribution to the proton-capture rate, which depends exponentially on the resonance energy. This state was first measured by Caggiano \textit{et al.}~at 1199(19)~keV via the \nuc{28}Si(\nuc{7}Li,\nuc{8}He)\nuc{27}P reaction \cite{CAGGIANO200164}. Subsequent Coulomb dissociation experiments at incident energies of 54.2 and 500 MeV/u, respectively, yielded energies of 1176(32)~keV \cite{TOGANO201184} and 1137(33)~keV \cite{MARGANIEC201693} for the $3/2^+$ resonance. The highest-precision measurements of the energy of the $3/2^+$ resonance are 1120(8)~keV from in-beam $\gamma$-ray spectroscopy of \nuc{27}P produced from one-proton knockout \cite{GADE2008044306} and 1125(2)~keV from $\beta$-delayed $\gamma$ rays observed in the recent $\beta$-decay spectroscopy of \nuc{27}S \cite{SUN1,SUN2}. 

Closer to stability, levels in \nuc{28}P have been identified using the reactions \nuc{28}Si(\textit{p},\textit{n}+$\gamma$)\nuc{28}P \cite{MOSS1972179,MIEHE197715} and \nuc{28}Si(\nuc{3}He,t)\nuc{28}P \cite{RAMSTEIN1979317}. Above the proton separation energy, $\gamma$ decay has been detected only from the 2104-keV state \cite{MOSS1972179,MIEHE197715}. However, in Ref.~\cite{MOSS1972179}, a 2216-keV $\gamma$ ray in coincidence with neutrons from the reaction of 23 MeV protons on a natural Si target was observed but with unknown origin. Later, a state in \nuc{28}P at 2216~keV was found in \nuc{28}Si(\nuc{3}He,t)\nuc{28}P \cite{RAMSTEIN1979317}. Furthermore, while upper limits for the lifetimes of two levels have been extracted using the Doppler-shift attenuation method \cite{MIEHE197715}, no experimental information is available for the lifetimes of the first two excited states in \nuc{28}P.
 
In this work, we report the observation of $\gamma$ rays in \nuc{26,27,28}P from in-beam $\gamma$-ray spectroscopy with both high-efficiency and high-resolution detector arrays. In \nuc{26}P, $\gamma$-ray transitions from a newly-found excited state above the 164.4-keV level have been detected. In \nuc{27}P, we present two new, high-precision measurements of the energy of the $3/2^+$ resonance important for the \nuc{26}Si(\textit{p},$\gamma$)\nuc{27}P reaction rate. In \nuc{28}P, we confirm the 2216-keV $\gamma$ ray and report a new $\gamma$ ray originating from the decay of a proton-unbound state. In addition, using the $\gamma$-ray lineshape method, the lifetimes of the first two excited states in \nuc{28}P have been measured for the first time. Experimental results are compared to calculations accounting for the onset of the Thomas-Ehrman effect \cite{EHRMAN195181,THOMAS195288} as the proton drip line is approached.

\section{Experiment}
As described in Ref.~\cite{LONGFELLOW201897}, three separate experiments were performed at the National Superconducting Cyclotron Laboratory \cite{GADE2016053003} to produce the data sets presented in this work. In the most recent of these measurements, excited states in \nuc{26}P and \nuc{27}P were populated in the reactions \nuc{9}Be(\nuc{26}Si,\nuc{26}P+$\gamma$)X and \nuc{9}Be(\nuc{26}Si,\nuc{27}P+$\gamma$)X, respectively. A secondary beam cocktail including \nuc{26}Si was produced from fragmentation of a 150~MeV/u \nuc{36}Ar primary beam on a 550 mg/cm$^2$ \nuc{9}Be target at the midacceptance position of the A1900 fragment separator \cite{MORRISSEY200390} with a 250~mg/cm$^2$-thick achromatic Al wedge degrader used for beam purification.

The secondary beam, which contained 14\% \textsuperscript{26}Si at 118 MeV/u, was impinged on a \nuc{9}Be secondary target with 287(3) mg/cm\textsuperscript{2} areal density placed at the reaction target position of the S800 spectrograph \cite{BAZIN2003629}. In order to detect de-excitation $\gamma$ rays, the secondary target was surrounded by CAESAR \cite{WEISSHAAR2010615}, a high-efficiency array of 192 closely-packed CsI(Na) scintillation crystals covering nearly 4$\pi$. The spatial granularity of CAESAR allowed for event-by-event Doppler reconstruction of the $\gamma$ rays emitted in-flight by reaction residues.

\begin{figure}
        \begin{center}
            \pic{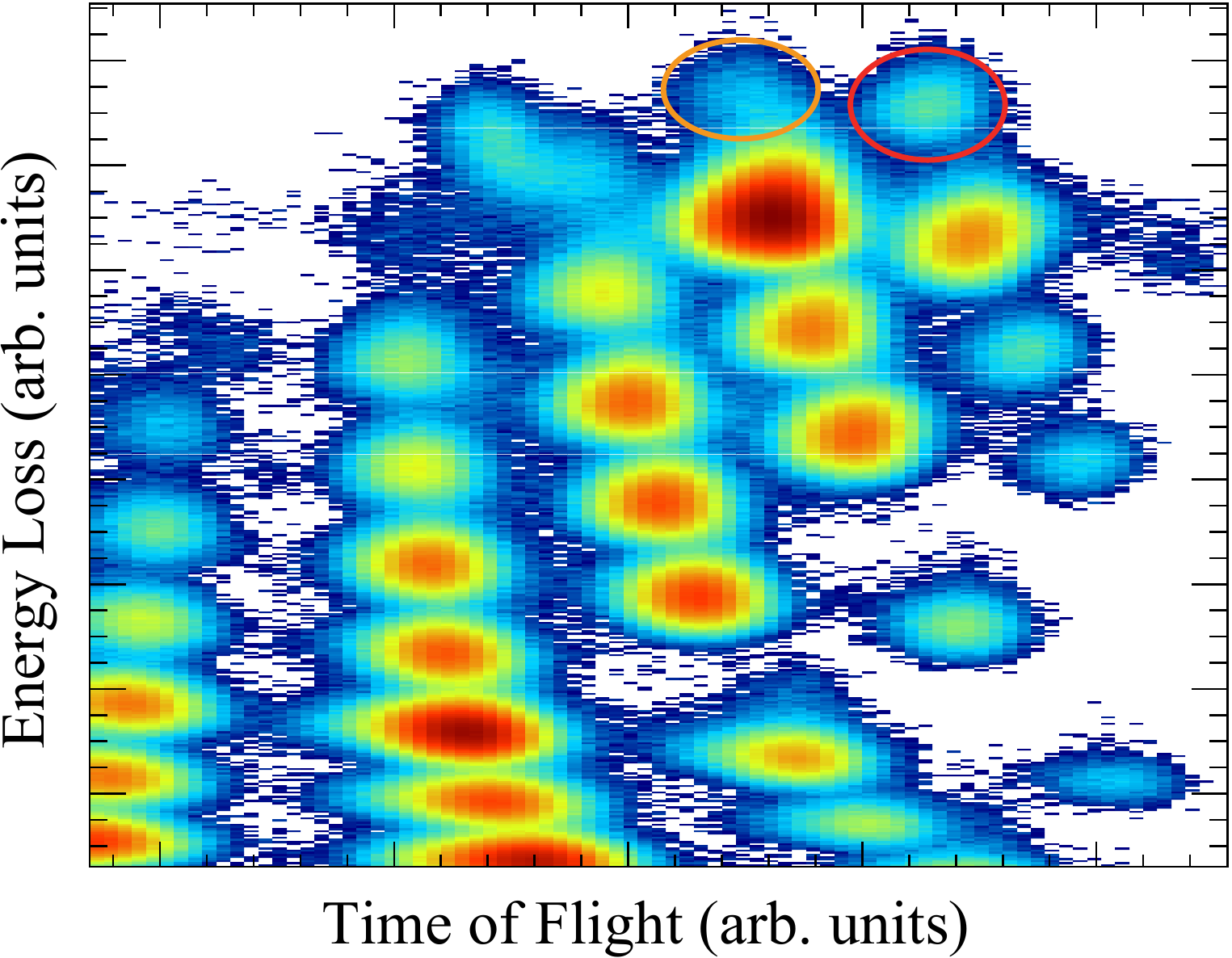}
            \caption{Particle identification plot for reaction residues produced from \nuc{26}Si projectiles impinged on a \nuc{9}Be target for the experiment using CAESAR for $\gamma$-ray detection (also provided in Fig.~1 of Ref.~\cite{LONGFELLOW201897}). The S800 ionization chamber was used to measure the energy loss and plastic scintillators in the beam line and the back of the S800 focal plane were used to measure the time of flight. \nuc{26}P (red oval) and \nuc{27}P (orange oval) reaction products are cleanly separated and identifiable.
            }
            \label{fig:pid}
        \end{center}
\end{figure}

Plastic timing scintillators in the beam line and the standard set of S800 focal-plane detectors \cite{YURKON1999291} were used to identify both the incoming projectiles and outgoing reaction products on an event-by-event basis. The energy loss versus time of flight particle identification plot for reaction residues from the \nuc{26}Si incoming beam is shown in Fig.~\ref{fig:pid}. For this measurement, the magnetic rigidity of the S800 spectrograph was set to center the \nuc{24}Si products from two-neutron knockout in the focal plane but the large acceptance of the S800 also allowed detection of \nuc{26,27}P residues.

The in-beam response of CAESAR after Doppler reconstruction was modeled using GEANT4 simulations benchmarked against the laboratory-frame energy spectra of various standard calibration sources. The 4.5-keV standard deviation for energies measured by CAESAR in-beam compared to energies of known $\gamma$-ray transitions in \nuc{22,23}Mg (see Fig.~2 of Ref.~\cite{LONGFELLOW201897}) was adopted as the systematic uncertainty in the Doppler-corrected energy and added in quadrature to the fit uncertainties. For the spectroscopy of \nuc{26,27}P, a nearest-neighbor addback procedure was used.
 
In addition to the data taken with CAESAR, two sets of high-resolution $\gamma$-ray data taken using SeGA \cite{MUELLER2001492}, an array of 32-fold segmented HPGe detectors, were analyzed to study the neutron-deficient phosphorus isotopes. The experimental schemes for both measurements were the same as described above except SeGA was used for $\gamma$-ray detection instead of CAESAR. The full technical details are available in Refs.~\cite{GADE2008044306} and \cite{YONEDA2006021303,REYNOLDS2010067303} with additional discussion in Ref.~\cite{LONGFELLOW201897}. For both experiments, SeGA was arranged in two rings with central angles of 90$\degree$ and 37$\degree$ with respect to the beam axis. In Ref.~\cite{GADE2008044306}, nine SeGA detectors were located at 90$\degree$ and seven occupied the 37$\degree$ ring while in Refs.~\cite{YONEDA2006021303,REYNOLDS2010067303}, ten SeGA detectors populated the 90$\degree$ ring with an additional seven placed in the 37$\degree$ ring. Doppler correction for each event was performed using the coordinates of the segment with the highest energy deposition.
 
In Ref.~\cite{GADE2008044306}, the secondary cocktail beam included \nuc{28}S and its isotone \nuc{27}P and excited states in the exotic nucleus \nuc{26}P were populated in the reactions \nuc{9}Be(\nuc{27}P,\nuc{26}P+$\gamma$)X and \nuc{9}Be(\nuc{28}S,\nuc{26}P+$\gamma$)X.  Furthermore, in different parts of the experiment described in Refs.~\cite{YONEDA2006021303,REYNOLDS2010067303}, secondary beams including \nuc{30}S, \nuc{29}P and \nuc{34}Ar, \nuc{33}Cl were utilized, allowing \nuc{27,28}P residues produced from several reactions to be identified using the S800 spectrograph.

\section{Results}
\begin{figure}
        \begin{center}
            \pic{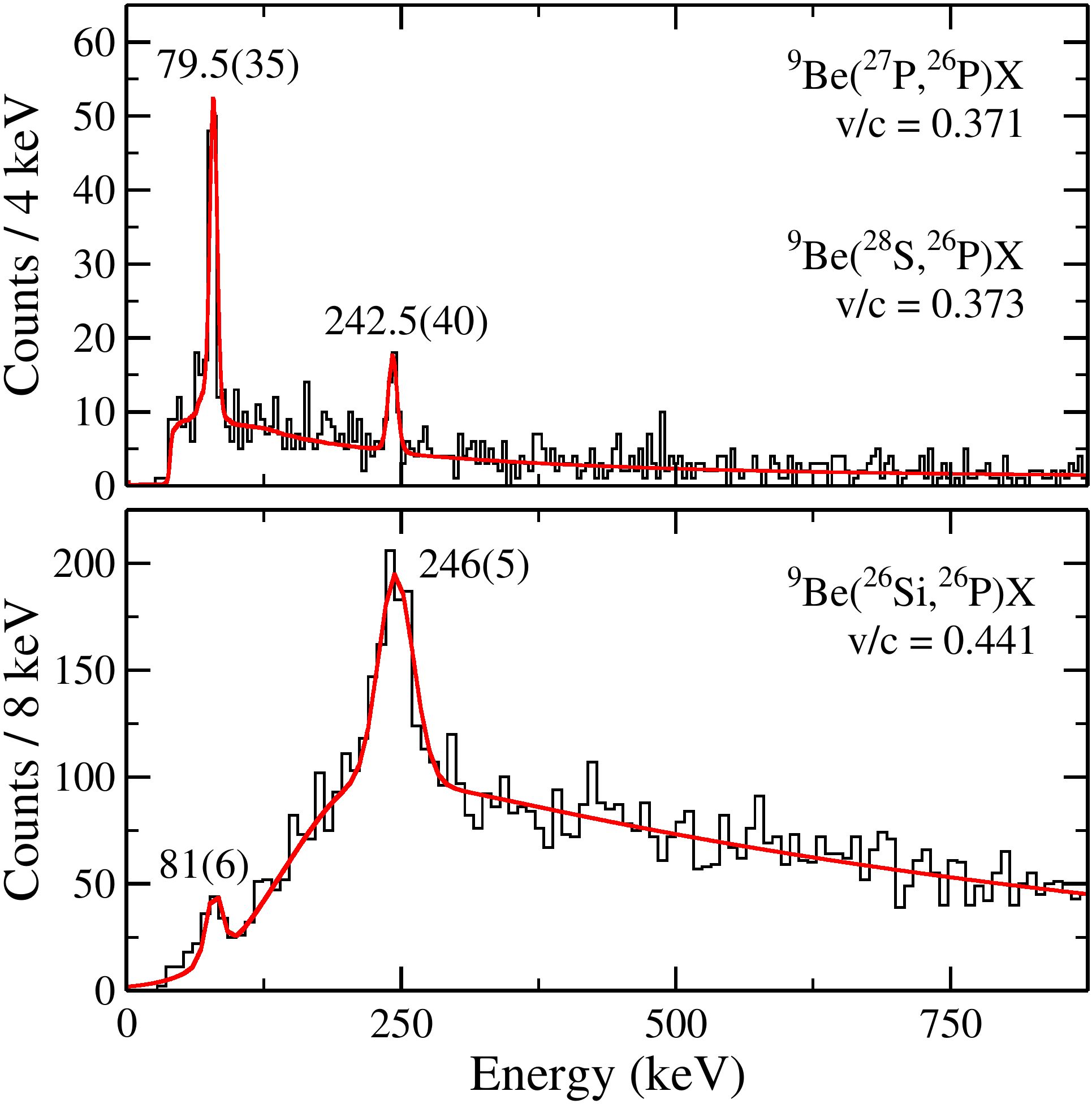}
            \caption{ 
                Doppler-corrected $\gamma$-ray spectra for \nuc{26}P measured with SeGA (top) from the experiment described in Ref.~\cite{GADE2008044306} and CAESAR (bottom). In the SeGA spectrum data from the two reaction channels specified have been added together. The red curves are GEANT4 simulations of the observed peaks with a double exponential background and a low-energy cutoff to account for detector thresholds.
            }
            \label{fig:p26}
        \end{center}
\end{figure}

The Doppler-corrected $\gamma$-ray spectra for \nuc{26}P measured with SeGA in the experiment described in Ref.~\cite{GADE2008044306} and with CAESAR are provided in Fig.~\ref{fig:p26}. The uncertainty-weighted average energies of the two peaks visible in both data sets are 80(3)~keV and 244(3)~keV. The 164(3)-keV difference between the observed transitions is in good agreement with the previously reported energies of 164.4(1)~keV \cite{NISHIMURA201466} and 164.4(3(2))~keV \cite{PEREZ201796} for the first excited state in \nuc{26}P, suggesting a level at 244~keV with $\gamma$-decay branches to the (3$^{+}$) ground state and the (1$^{+}$) first excited state. Since the 164.4-keV level has a half-life on the order of 100~ns and the velocity of the \nuc{26}P projectiles was around 0.4c, the $\gamma$ decay of this state occurred far behind the target on average and could not be measured in the experiments discussed here. 

After correcting for the energy-dependent $\gamma$-ray detection efficiencies and low-energy detection thresholds for the arrays, the branching ratios for the $\gamma$ decay of the 244-keV state were calculated. For the SeGA data, the relative $\gamma$-decay intensity is larger to the first excited state (100(13)\%) than to the ground state (49(13)\%). Although the peak area for the 80-keV transition in the CAESAR data is comparatively small, accounting for detector thresholds gives results consistent with the SeGA data: 100(21)\% normalized intensity for the decay to the first excited state compared to 39(21)\% for the decay to the ground state. The combined results for both data sets are 100(11)\% relative branching to the first excited state compared to 46(11)\% to the ground state. The uncertainties for the reported branching ratios include statistical, fit, and efficiency uncertainties added in quadrature. The branching ratios obtained using the CAESAR spectra with and without addback were in agreement.

In order to tentatively assign a spin-parity to the 244-keV level in \nuc{26}P, comparisons were made with the experimental information available on the mirror nucleus, \nuc{26}Na, and with the results of shell-model calculations. In the mirror nucleus, the second excited state is at 232.7(6)~keV and has a spin-parity of 2$^{+}$. Similarly to the observed 244-keV state in \nuc{26}P, the 232.7(6)-keV level in \nuc{26}Na $\gamma$ decays to the 82.2(6)-keV 1$^{+}$ first excited state and the 3$^{+}$ ground state with normalized intensities of 75.1\% and 100\% via mixed $M1/E2$ transitions \cite{ENSDF}. The next excited state, which also has spin-parity 2$^{+}$, is at 406.7(5)~keV and $\gamma$ decays mainly to the ground state (100\%) with smaller branches to the 1$^+$ excited state (14.6\%) and to the first 2$^+$ excited state (2.8\%) \cite{ENSDF}. Shell-model calculations were performed in the \textit{sd} model space using the USDA and USDB Hamiltonians \cite{BROWN200674} with added charge-dependent (CD) parts derived from fits to isobaric mass multiplet data \cite{ORMAND1989491}. USDA+CD and USDB+CD both predict a 3$^{+}$ ground state and a low-lying 1$^{+}$ excited state at 32 and 109~keV, respectively. Using the shell-model $B(E2)$ values and the experimental level energy of 164.4~keV, the half-life of the 1$^{+}$ state is predicted to be 128~ns for USDA+CD and 174~ns for USDB+CD, in broad agreement with the experimental values of 120(9)~ns \cite{NISHIMURA201466} and 104(14)~ns \cite{PEREZ201796}. In both shell-model predictions for \nuc{26}P, the second and third excited states have spin-parity 2$^{+}$, like in the mirror, and are under 500~keV in excitation energy while the next excited state is above 1.2~MeV. The first 2$^{+}$ (164~keV) decays only 3\% of the time to the first excited state relative to the ground state (100\%) in USDA+CD and has a lifetime of 114~ps, while in USDB+CD the first 2$^{+}$ (123~keV) has a 137-ps lifetime and $\gamma$ decays to the first excited state with 32\% intensity relative to the ground state (100\%). Using the experimental values for the energies (164.4 and 244~keV) and the shell-model values for the transition strengths (predominantly M1), the lifetimes for the first 2$^+$ are lowered to 35~ps in USDA+CD and 22~ps in USDB+CD. In both USDA+CD and USDB+CD, the second 2$^{+}$ excited states (423 and 332~keV) have lifetimes of around 1~ps and decay mostly to the ground state (100\%) with the remaining relative strength split between the 1$^+$ (19\% in USDA+CD, 47\% in USDB+CD) and 2$^+_1$ (15\% in USDA+CD, 14\% in USDB+CD) excited states. There is no clear evidence for additional $\gamma$-ray transitions in Fig.~\ref{fig:p26}. The lifetime of the experimentally observed state at 244~keV in \nuc{26}P could not be determined using the $\gamma$-ray lineshape method due to limited statistics and the low-energy threshold cutting into the Doppler-reconstructed 80-keV peak for detectors in the 90$\degree$ ring. 

Based on the arguments above, we assign the newly-observed 244-keV level in \nuc{26}P a tentative spin-parity of (2$^{+}$). Estimates for the unmeasured proton separation energy in \nuc{26}P include 140(200)~keV \cite{WANG201236}, 0(90)~keV \cite{THOMAS200421}, $-119(16)$~keV \cite{TIAN201387}, and 85(30)~keV \cite{FORTUNE201796}. In the experimental setup discussed here, any proton-decay branch from the newly-observed 244-keV level could not have been detected. Furthermore, the experiment did not allow for discrimination between the direct population of the isomeric state and the ground state, precluding the extraction of individual cross sections. According to USDB+CD, the spectroscopic factors for one-neutron knockout from the $1/2^+$ ground state of \nuc{27}P are 1.665 to the $3^+$ ground state, 0.058 to the $1^+$ isomeric state, 0.648 to the first $2^+$ state, and 0.717 to the second $2^+$ state of \nuc{26}P. If these spectroscopic factors are accurate, then the absence of other $\gamma$-ray transitions in the one-neutron knockout data in the top panel of Fig.~\ref{fig:p26} suggests that the energy of the second $2^+$ level is high enough above the separation energy that proton decay is its dominant decay mode. 

\begin{figure}
        \begin{center}
            \pic{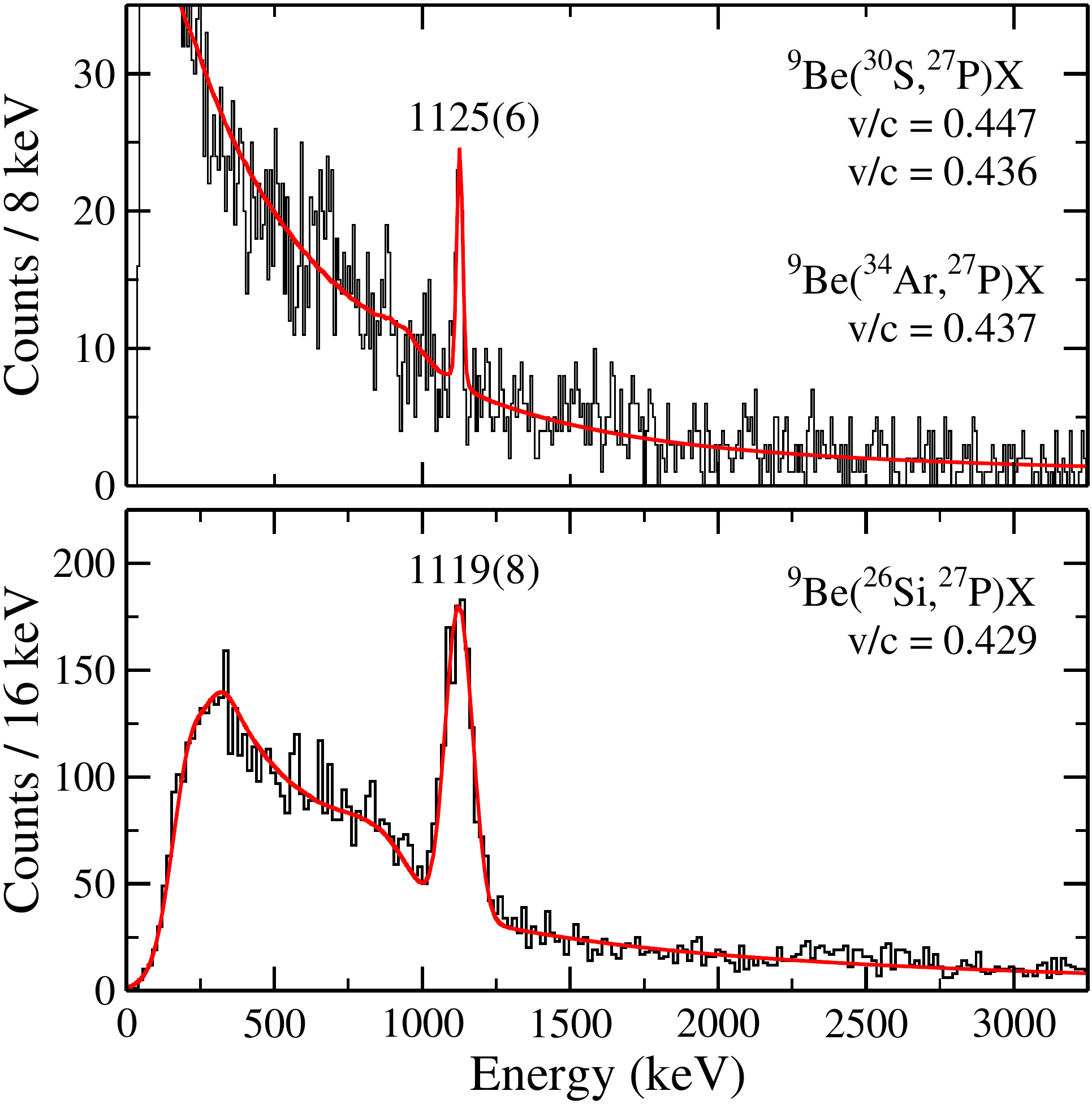}
            \caption{ 
                Doppler-corrected $\gamma$-ray spectra for \nuc{27}P measured with SeGA (top) from the experiment described in Refs.~\cite{YONEDA2006021303,REYNOLDS2010067303} and CAESAR (bottom). The SeGA spectrum includes data from \nuc{9}Be(\nuc{30}S,\nuc{27}P+$\gamma$)X taken under two S800 magnetic field settings and \nuc{9}Be(\nuc{34}Ar,\nuc{27}P+$\gamma$)X. The transition observed with both arrays is consistent with the $\gamma$ decay of the $3/2^+$ state reported at 1120(8)~keV from the previous in-beam $\gamma$-ray spectroscopy experiment \cite{GADE2008044306} and at 1125(2)~keV in a recent \nuc{27}S $\beta$-decay measurement \cite{SUN1,SUN2}.
            }
            \label{fig:p27}
        \end{center}
\end{figure}

Displayed in Fig.~\ref{fig:p27} are the Doppler-corrected $\gamma$-ray spectra for \nuc{27}P obtained from the CAESAR data and the SeGA data from the experiment detailed in Refs.~\cite{YONEDA2006021303,REYNOLDS2010067303}. The energies of the peaks visible in both spectra are 1125(6)~keV for SeGA and 1119(8)~keV for CAESAR. Although CAESAR has modest resolution conpared to SeGA, the uncertainties for the reported energies are similar due to the statsitics advantage of the high-efficiency array. Both values agree well with the highest-precision measurements available for the energy of the astrophysically-important $3/2^+$ resonance, 1120(8)~keV from in-beam $\gamma$-ray spectroscopy \cite{GADE2008044306} and 1125(2)~keV from the detection of $\beta$-delayed $\gamma$ rays following the decay of \nuc{27}S \cite{SUN1,SUN2}. The proton separation energy for \nuc{27}P deduced in Refs.~\cite{SUN1,SUN2} from the measured $\gamma$ and proton decays of the $3/2^+$ resonance is 807(9)~keV and no $\gamma$ decay branch from the $5/2^+$ second excited state at 1569(12)~keV was observed. In this work, there are also no clear $\gamma$-ray transitions corresponding to the decay of the $5/2^+$ or higher-lying resonances.   

\begin{figure}
        \begin{center}
            \pic{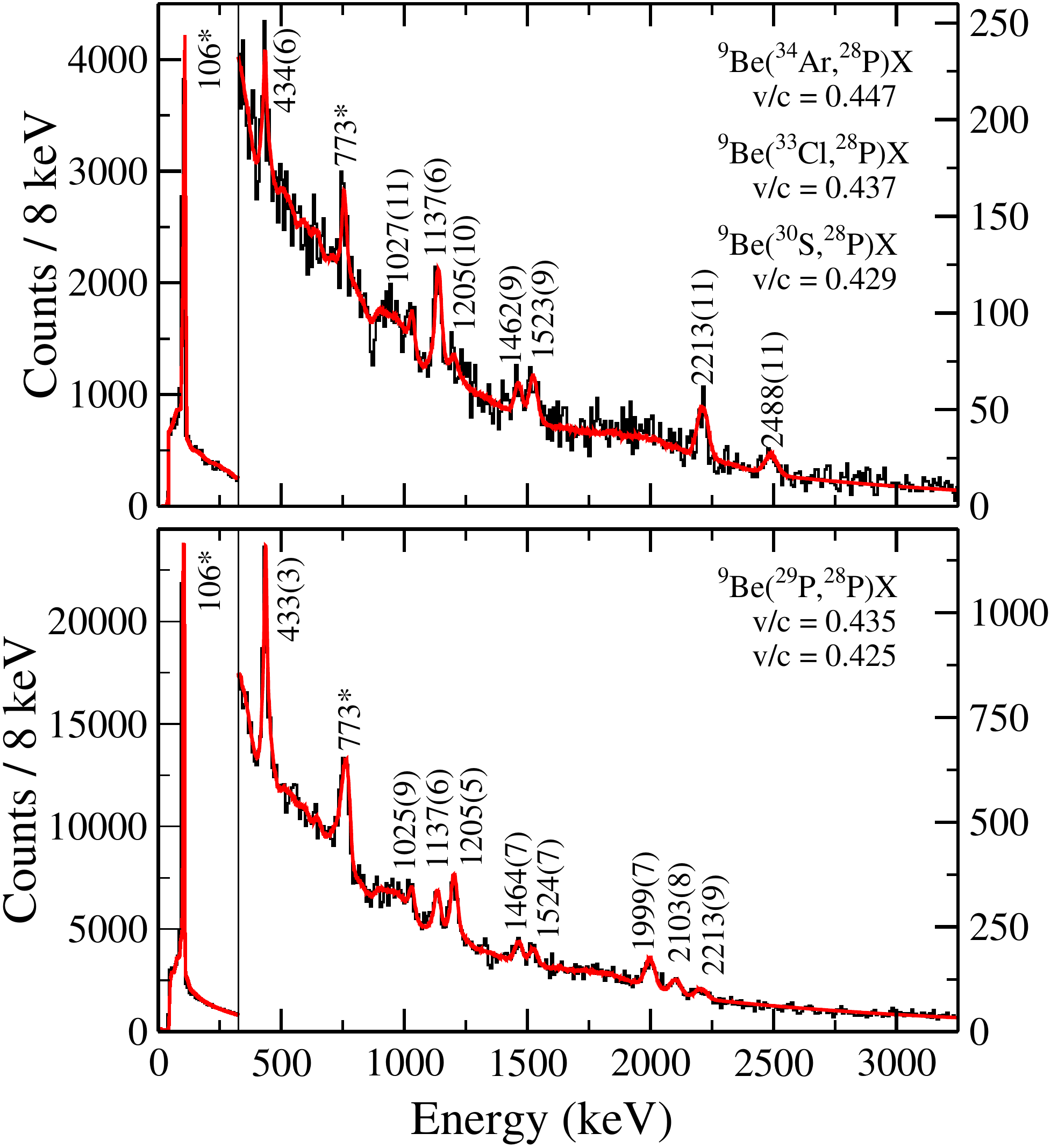}
            \caption{ 
                Doppler-corrected $\gamma$-ray spectra for \nuc{28}P from multi-nucleon removal from \nuc{34}Ar, \nuc{33}Cl, and \nuc{30}S (top) and one-neutron knockout using two magnetic-rigidity settings of the S800 spectrograph (bottom) measured with SeGA in the experiment described in Refs.~\cite{YONEDA2006021303,REYNOLDS2010067303}. The transitions labeled 106* and 773* originate from levels with appreciable lifetimes (see Fig.~\ref{fig:p28_tau}).
            }
            \label{fig:p28}
        \end{center}
\end{figure}

In the experiment described in Refs.~\cite{YONEDA2006021303,REYNOLDS2010067303}, \nuc{28}P was produced in several reactions. The top panel of Fig.~\ref{fig:p28} shows the combined Doppler-corrected $\gamma$-ray spectrum for \nuc{28}P produced from the removal of multiple nucleons from \nuc{34}Ar, \nuc{33}Cl, and \nuc{30}S while the bottom panel shows \nuc{28}P produced from one-neutron knockout under two magnetic field settings of the S800 spectrograph. The energies for all observed $\gamma$ rays below 2.2~MeV agree within uncertainties with previously reported transitions. The 2213(7)-keV and 2488(11)-keV transitions, which are slightly above the proton separation energy of 2052.3(12)~keV, are not given in the compilation of Ref.~\cite{ENSDF}. However, an unplaced 2216-keV $\gamma$ ray in coincidence with neutrons from the reaction of 23 MeV protons on a natural Si target was reported in Ref.~\cite{MOSS1972179} and the energies of these newly-observed peaks match well with the energies of the 2216(5)-keV and 2483(5)-keV states measured using \nuc{28}Si(\nuc{3}He,t)\nuc{28}P \cite{RAMSTEIN1979317}. Therefore, we interpret them as the ground-state $\gamma$ decays of these levels.

Based on comparisons with shell-model calculations and the mirror nucleus, \nuc{28}Al, the 2216-keV state, which has previously been assigned a tentative spin-parity of (3$^{+}$,4$^{+}$) with (4$^{+}$) favored \cite{RAMSTEIN1979317} is interpreted as the (4$^{+}$). In both \nuc{28}Na and the shell model, the first 4$^{+}$ state, which is at 2271.745(19)~keV in \nuc{28}Al \cite{ENSDF} and 2221~keV in USDB+CD, decays predominantly to the 3$^{+}$ ground state while the first excited 3$^{+}$ states in the mirror and in the shell model are close to 3~MeV and have large branching ratios to several levels. Futhermore, the observed predominant $\gamma$ decay of the 2483-keV level to the ground state supports the (5$^{+}$) assignment given in Ref.~\cite{RAMSTEIN1979317}.

\begin{figure}
        \begin{center}
            \pic{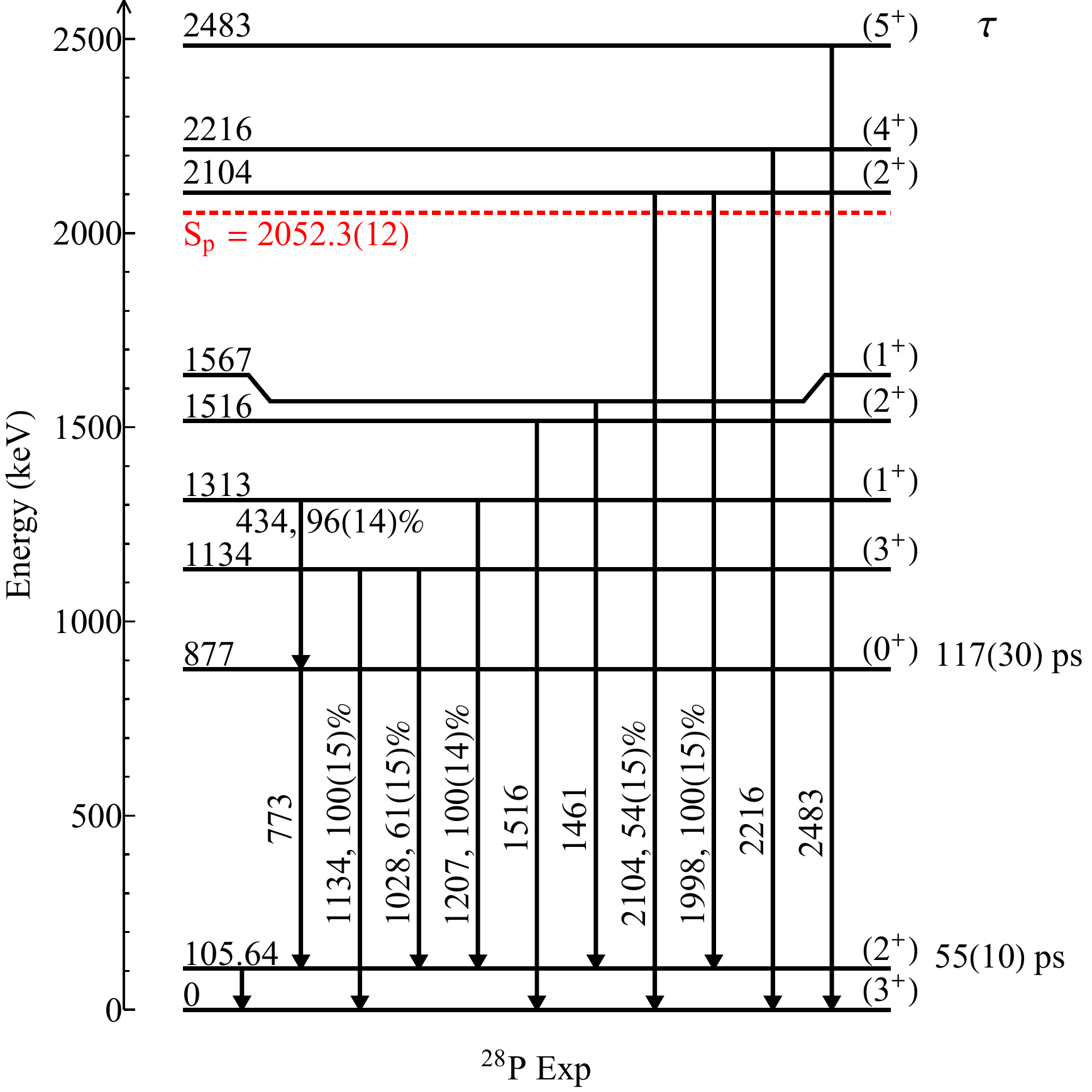}
            \caption{ 
                \nuc{28}P level scheme including the $\gamma$-ray transitions observed in this work. The energies for transitions and states are taken from Ref.~\cite{ENSDF}. The 2216-keV and 2483-keV transitions have been assigned as the ground-state decays of known states. The placement of the other transitions is the same as in previous experiments. 
            }
            \label{fig:p28_levels}
        \end{center}
\end{figure}

The observed transitions are summarized in the level scheme given in Fig.~\ref{fig:p28_levels}. Energy differences and information from $\gamma\gamma$ coincidences were used to construct the level scheme, which is in agreement with previous results \cite{ENSDF}. The normalized branching ratios for the 1028-keV and 1134-keV $\gamma$ decays from the 1134-keV state extracted in this work are 61(15)\% and 100(15)\%, which are in agreement with the known values of 89(16)\% and 100(16)\% \cite{MOSS1972179} and 45(23)\% and 100(23)\% \cite{MIEHE197715} within uncertainties. For the 434-keV and 1207-keV transitions from the 1313-keV level, relative branching ratios of 96(14)\% and 100(14)\% were measured, compared to the previous values of 100(37)\% and 85(37)\% \cite{MIEHE197715}. Finally, the relative intensities for the 1998-keV and 2104-keV $\gamma$ decays from the 2104-keV level were found to be 100(15)\% and 54(15)\%, which are similar to the 100(24)\% and 69(24)\% \cite{MOSS1972179} and 100(26)\% and 64(26)\% \cite{MIEHE197715} relative intensities available in the literature. 

The lifetimes of the first two excited states in \nuc{28}P are long enough to be measured using the $\gamma$-ray lineshape method, as has been done with SeGA arranged in 90$\degree$ and 37$\degree$ rings in previous works, e.g.~Refs.~\cite{TERRY200877,LEMASSON201285,BAUGHER201286,STROBERG201286,MORSE201490,
NICHOLS201591,BAUGHER201693,MILNE201693}. At the beam velocities discussed in this work, a state with lifetime on the order of 100~ps will $\gamma$ decay around 1 cm downstream of the target on average, causing the Doppler-corrected peak to have a low-energy tail and a centroid below the correct energy if the mid-target position and mid-target velocity are used to calculate the angle of the $\gamma$ ray relative to the beam axis. Therefore, the lifetime of the state can be extracted by fitting GEANT4 simulations with different lifetimes to the asymmetric lineshape of the peak observed experimentally.

The v/c values used in the Doppler reconstruction were determined from measuring the parallel momentum distribution in the S800 spectrograph and extrapolating back to the mid-target velocity using the ATIMA stopping powers \cite{ATIMA} implemented in LISE++ \cite{LISE++}. While the geometric description of the array is accurate to the order of millimeters, the sensitivity of Doppler reconstruction to detector position could be exploited to further refine the detector locations. The positions of the SeGA detectors along the beam axis with respect to the target were adjusted individually by aligning high-statistics peaks with known short lifetimes to their known Doppler-corrected energies. As a test of the method, the lifetime of the 780.8-keV level in \nuc{27}Si, which was determined previously to be 50(6)~ps using the recoil-distance method \cite{MAURITZSON1971174}, was measured. 

\begin{figure}
        \begin{center}
            \pic{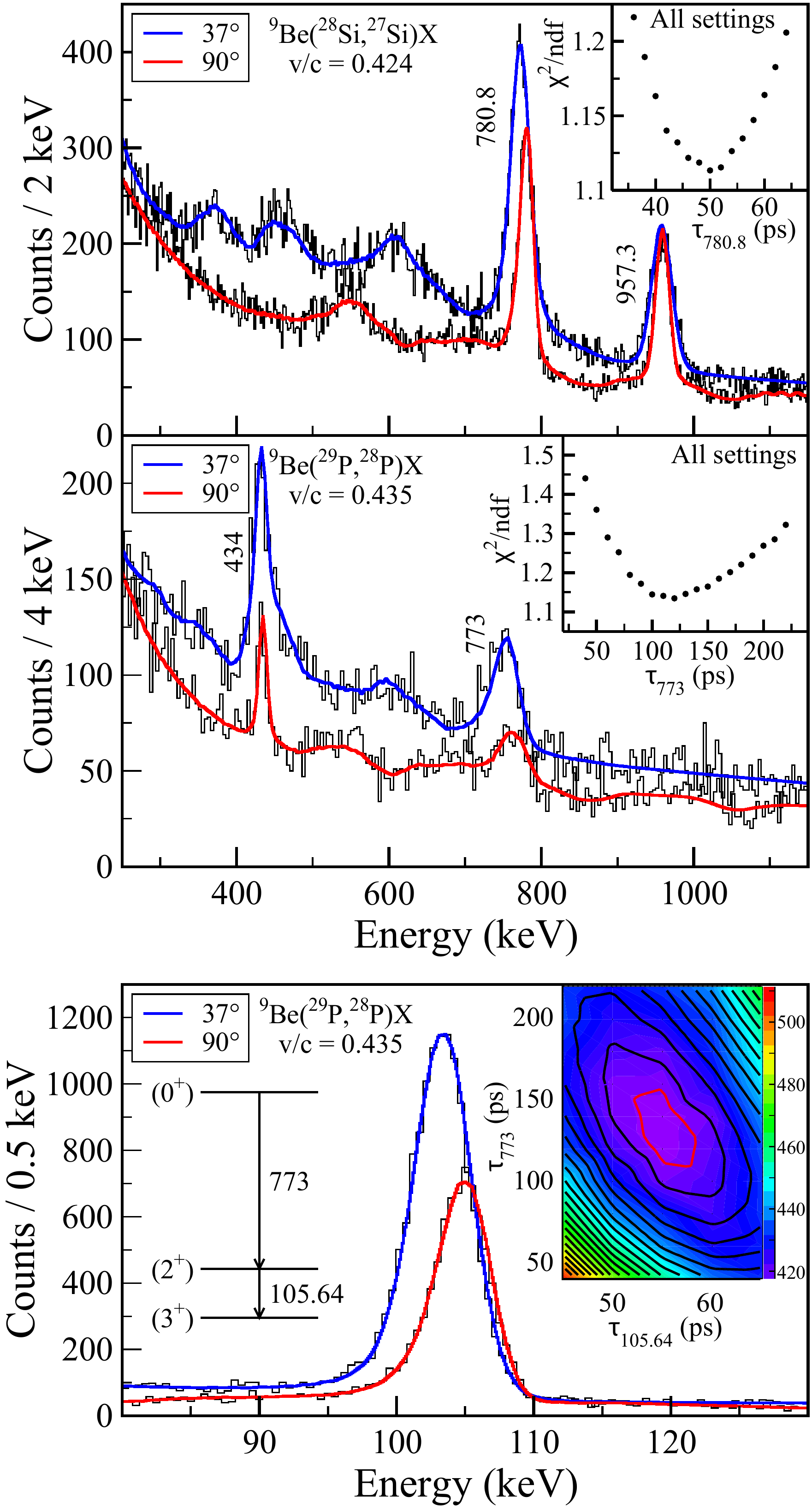}
            \caption{ 
                Doppler-corrected $\gamma$-ray spectra showing the effect of level lifetimes on the 780.8-keV transition in \nuc{27}Si (top), the 773-keV transition in \nuc{28}P (middle), and the 105.64-keV transition in \nuc{28}P (bottom) as measured with SeGA in the experiment described in Refs.~\cite{YONEDA2006021303,REYNOLDS2010067303} for data from the S800 spectrograph magnetic-rigidity setting centered on \nuc{9}Be(\nuc{30}S,\nuc{29}S+$\gamma$)X. The insets show the combined $\chi^2$ distributions of the lifetime fits for the 37$\degree$ and 90$\degree$ data for both settings, \nuc{9}Be(\nuc{30}S,\nuc{29}S+$\gamma$)X and \nuc{9}Be(\nuc{30}S,\nuc{28}S+$\gamma$)X.
            }
            \label{fig:p28_tau}
        \end{center}
\end{figure}

In this work, \nuc{27}Si produced from one-neutron knockout was studied using two magnetic-rigidity settings of the S800 spectrograph, one centered on \nuc{9}Be(\nuc{30}S,\nuc{29}S+$\gamma$)X and the other centered on \nuc{9}Be(\nuc{30}S,\nuc{28}S+$\gamma$)X. The top panel of Fig.~\ref{fig:p28_tau} shows the $\gamma$-ray spectra for \nuc{27}Si for the \nuc{9}Be(\nuc{30}S,\nuc{29}S+$\gamma$)X setting with GEANT4 simulations of the observed peaks. Laboratory-frame background lines from electron-positron annihilation and neutron-induced reactions in the Ge detectors and Al beam pipe, which have been smeared due to the Doppler correction of the in-flight $\gamma$ rays, were also included. The effect of the lifetime of the 780.8-keV state on the 37$\degree$ and 90$\degree$ spectra is clear in juxtaposition with the 957.3-keV transition, which originates from a level with a comparatively short lifetime of 1.73(12)~ps \cite{ENSDF}. The best fit for the lifetime of the 780.8-keV level was found using $\chi^2$ minimization. The fit region used was from 650 to 850~keV with the exponential background and contribution of other transitions fixed from fits in the range shown in the top panel of Fig.~\ref{fig:p28_tau}. Only the scaling factor of the GEANT4 simulation for each lifetime was allowed to vary. The inset of the top panel in Fig.~\ref{fig:p28_tau} shows the reduced $\chi^2$ plot for all \nuc{27}Si data (37$\degree$ and 90$\degree$ rings for both S800 spectrograph magnetic-rigidity settings). Results for both rings in both settings were consistent within uncertainties.

The statistical uncertainty for the lifetime derived from all data was taken from the 95\% confidence interval ($\chi_{\text{min}}^2$+4) to be 3~ps. In order to estimate systematic uncertainties, the effects of varying the v/c used in Doppler reconstruction and the offsets of the SeGA detectors on the extracted lifetime of the 780.8-keV state were explored. The offsets used in the geometry of the setup and the choice of v/c provided 8\% and 4\% uncertainty, respectively. Smaller systematic uncertainties (less than 2\% each) included the choice of bin size, the functional form used to determine the minimum $\chi^2$ value, the fit range and background used, and indirect feeding of the 780.8-keV level. Adding all statistical and systematic uncertainties in quadrature, we report a lifetime of 51(6)~ps for the 780.8-keV state in \nuc{27}Si, in excellent agreement with the previous value of 50(6)~ps \cite{MAURITZSON1971174}.

The same procedure was then applied to the 773(3)-keV transition from the 877(2)-keV state to the 105.64(10)-keV level in \nuc{28}P. The energies and uncertainties for the transition and levels were taken from the literature values \cite{ENSDF}. The middle panel of Fig.~\ref{fig:p28_tau} shows the 434-keV peak, which originates from a level with an expected lifetime of around 0.5~ps from shell-model calculations and comparison with the mirror nucleus, and the 773-keV peak, which is not aligned in the 37$\degree$ and 90$\degree$ detectors due to the long lifetime of the 877-keV state. The best-fit lifetimes for the 37$\degree$ and 90$\degree$ data sets taken with the two different magnetic rigidities of the S800 spectrograph were in agreement within uncertainties. Using the statistical uncertainty of $^{+28}_{-24}$~ps derived from the fit to the summed $\chi^2$ plot shown in the inset and the same systematic uncertainties as for \nuc{27}Si, the lifetime of the 877-keV level is 117(30)~ps, where the lower bound for the uncertainty was increased since the 877-keV level is fed strongly by the 434-keV transition.

As visible in the bottom panel of Fig.~\ref{fig:p28_tau}, the 773-keV transition feeds the long-lived 105.64-keV level. In order to determine the lifetime of this state, the peak corresponding to the 105.64-keV transition was fit with GEANT4 simulations that included feeding effects. Corrected for efficiency, the 105.64-keV state was fed 13(3)\% of the time by the 773-keV transition. The lifetimes of the 105.64-keV and 773-keV transitions were varied simultaneously in the fits. Other transitions to the 105.64-keV level were assumed to have negligible lifetimes, consistent with their alignment in the 37$\degree$ and 90$\degree$ detectors. The systematic uncertainty for the two extracted lifetimes was found from the extrema of the projections of the $\chi_{\text{min}}^2$+4 contour shown in red in the inset of Fig.~\ref{fig:p28_tau}. This plot contains the summed $\chi^2$ values of the fits to both the 37$\degree$ and 90$\degree$ detectors in both S800 spectrograph magnetic-rigidity settings with the fits performed between 80 and 130~keV. Furthermore, the feeding percentage was varied within its uncertainties to quantify its effect on the extracted lifetimes. From these results, an additional 15\% uncertainty was added in quadrature to the systematic uncertainties established from the \nuc{27}Si data. From these fits, the lifetime for the 105.64-keV state is 55(10)~ps and the lifetime for the 877-keV state is 132(35)~ps, in agreement with the result of the direct fit to the 773-keV transition, 117(30)~ps. We recommend adopting 117(30)~ps for the lifetime of the 877-keV state because the feeding percentage used in the simultaneous fit was determined by the direct fit to the 773-keV transition.  

The 780.8(2)-keV transition from the first 1/2$^+$ state to the 5/2$^+$ ground state in \nuc{27}Si is predominately of $E2$ character. Using a lifetime of 50.5(35)~ps, the $B(E2)$ strength is 55.5(40)~$e^2$fm$^4$. For this transition, USDB+CD predicts a $B(E2)$ of 81.0~$e^2$fm$^4$ giving a ratio of experiment to shell model of 0.69(5). In the mirror nucleus \nuc{27}Al, the $B(E2)$ for the analog transition is 38.0(11)~$e^2$fm$^4$ experimentally compared to 52.8~$e^2$fm$^4$ as calculated by USDB+CD for a ratio of 0.72(2). Although the predictions for the $B(E2)$ strengths are larger than the observed values, these ratios show that the proportionate change in $B(E2)$ from \nuc{27}Al to \nuc{27}Si is well-reproduced by the shell model.

In \nuc{28}P, the 105.64(10)-keV decay from the first (2$^+$) level to the 3$^+$ ground state is predicted to be $M1$ by shell-model calculations and in the mirror nucleus, the analog transition has $\delta=+0.001(6)$ \cite{ENSDF}. Assuming no mixing and using 55(10)~ps for the lifetime, the $B(M1)$ strength is 0.877(160)~$\mu_{N}^2$ compared to 0.688~$\mu_{N}^2$ as predicted by USDB+CD. In \nuc{28}Al, the experimental $B(M1)$ strength is 0.662(16)~$\mu_{N}^2$ while USDB+CD gives 0.546~$\mu_{N}^2$. In this case, the shell-model predictions for the strengths are smaller than the observed values. However, the experiment-to-shell-model ratio of 1.27(23) for \nuc{28}P is similar to the value of 1.21(3) for \nuc{28}Al.

Using 117(30)~ps for the lifetime of the $(0^+)$ state in \nuc{28}P at 877~keV, which decays to the (2$^+$) state at 105.64~keV via a 773(3)-keV $\gamma$-ray, the $B(E2)$ strength is 25.2(65)~$e^2$fm$^4$ while the USDB+CD prediction is 40.1~$e^2$fm$^4$ for a ratio of 0.63(16). In the mirror nucleus, the experimental and shell-model $B(E2)$ values are 23.1(14) and 29.3~$e^2$fm$^4$, respectively, giving a ratio of 0.79(5). Although the ratio for \nuc{28}Al is larger than for \nuc{28}P, the values are consistent within uncertainties and are similar to the ratios found for the $B(E2)$ strengths in \nuc{27}Si and \nuc{27}Al above.

\section{Discussion}
The lowering of the proton separation energy in nuclei as the proton drip line is approached can greatly affect nuclear structure properties. For weakly-bound protons in the 1$s_{1/2}$ orbital, the lack of angular momentum barrier causes an extension of the wavefunction in the radial direction compared to that of the mirror nucleus and leads to an asymmetry in the energies of analog states known as the Thomas-Ehrman shift \cite{EHRMAN195181,THOMAS195288}. For \nuc{26,27,28}P, shell-model calculations predict that the spectroscopic factors for the one proton 1$s_{1/2}$ overlaps with \nuc{25,26,27}Si, respectively, are larger for the ground state than for the excited states (see Table~\ref{tab:sf}). Here, we use shell-model spectroscopic factors to calculate theoretical Thomas-Ehrman shifts and compare the results to the observed values.

The predicted Thomas-Ehrman shift for the proton 1$s_{1/2}$ orbital as a function of the proton separation energy (S$_p$) was calculated using a Woods-Saxon potential with a \nuc{28}Si core. The Thomas-Ehrman shift was tabulated as the difference in proton and neutron 1$s_{1/2}$ single-particle energies relative to the difference in proton and neutron 0$d_{5/2}$ single-particle energies. The inital parameters for the Woods-Saxon potential were taken from Ref.~\cite{BOHRMOTTELSON}. The radius and potential depth of the central potential were then varied in order to reproduce the rms charge radii of \nuc{28}Si and \nuc{31}P.

For the central and spin-orbit potentials, r$_0$= 1.25~fm, r$_{so}$=1.27~fm, and a$_0$=a$_{so}$=0.67~fm were used. In the Coulomb term, the radius was taken as r$_c$=1.20~fm. The spin-orbit potential strength V$_{so}$ was fixed at 22~MeV. Using these parameters at a central potential depth of V$_0$=$-55.6$~MeV, the calculated rms charge radii for \nuc{28}Si and \nuc{31}P are 3.1408~fm and 3.1775~fm compared to the experimental values of 3.1224(24)~fm and 3.1889(19)~fm \cite{RADII}, respectively. In addition, the  proton separation energies calculated for \nuc{28}Si and \nuc{31}P, 11.863~MeV and 7.308~MeV, are in broad agreement with the literature values of 11.58502(10)~MeV and 7.29655(2)~MeV \cite{ENSDF}. 

\begin{figure}
        \begin{center}
            \pic{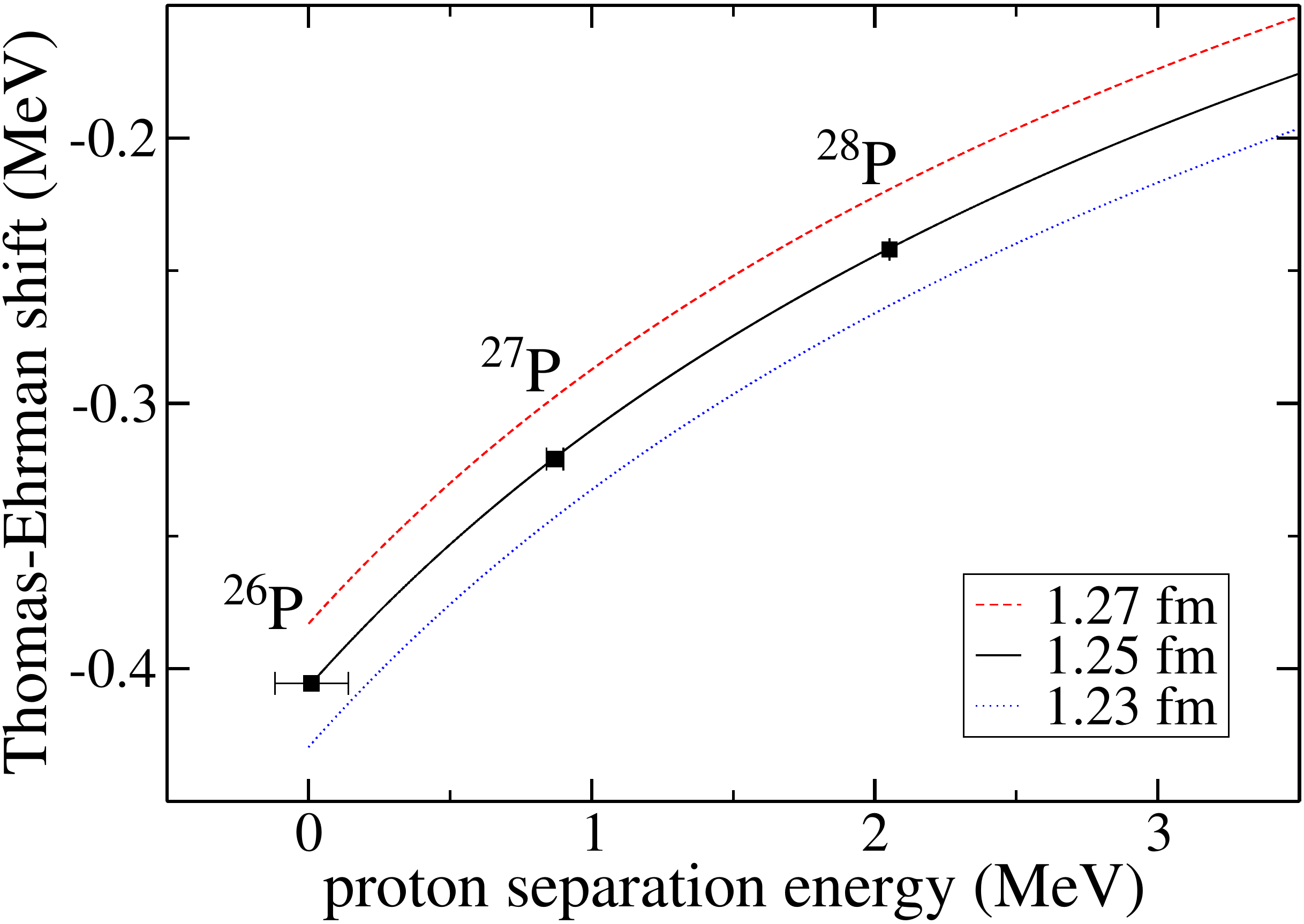}
            \caption{ 
                Thomas-Ehrman shift as a function of proton separation energy calculated using a Woods-Saxon potential with several values of $r_0$. The ground-state proton separation energies for \nuc{26,27,28}P are labeled. 
            }
            \label{fig:TE_sep}
        \end{center}
\end{figure}

As shown in Fig.~\ref{fig:TE_sep}, V$_0$ was varied to examine the Thomas-Ehrman shift at different values of the proton separation energy. Increasing r$_0$ by 0.02~fm causes the magnitude of the Thomas-Ehrman shift to increase by about 23~keV while decreasing r$_0$ by 0.02~fm decreases the magnitude of the Thomas-Ehrman shift by around 24~keV. 

Since the mass of \nuc{26}P has not been measured, the proton separation energy shown in Fig.~\ref{fig:TE_sep} and used in the calculations below for \nuc{26}P is 10.5~keV, the midpoint of the two extreme predictions of 140(200)~keV \cite{WANG201236} and $-119(16)$~keV \cite{TIAN201387}. The error bars in Fig.~\ref{fig:TE_sep} extend from $-119$ to 140~keV. The proton separation energies and uncertainties for \nuc{27,28}P were taken from Ref.~\cite{ENSDF}.

The calculated Thomas-Ehrman shifts relative to the ground state were then applied to the known experimental energies (E$_\text{mirror}$) for levels in the mirror nuclei, \nuc{26}Na, \nuc{27}Mg, and \nuc{28}Al, to determine the predicted energies of the states in \nuc{26,27,28}P (E$_\text{mirror+TE}$) using the equation:

\begin{equation}
\begin{split}
\text{E}_{\text{mirror}} + \text{TE}(\text{S}_p - \text{E}_{\text{mirror+TE}})\times\text{C}^2\text{S}(l=0,\text{E}^*) \\
- \text{TE}(\text{S}_p)\times\text{C}^2\text{S}(l=0,\text{gs}) = \text{E}_{\text{mirror+TE}}.
\end{split}
\end{equation} 
Values of the Thomas-Ehrman (TE) shift for unbound states were extrapolated from Fig.~\ref{fig:TE_sep}. Spectroscopic factors for the proton 1$s_{1/2}$ overlaps with states in the Si isotopes for the P ground states (C$^2$S($l=0$,gs)) and excited states (C$^2$S($l=0$,E$^*$)) were calculated using the USDB interaction. USDB was used rather than USDB+CD since the charge-dependent part already includes some Thomas-Ehrman shift. The spectroscopic factors calculated using USDA were comparable. For \nuc{26,27,28}P, the sum of spectroscopic factors for the proton 1$s_{1/2}$ overlap with all shell-model levels under 4 MeV in excitation energy in \nuc{25,26,27}Si, respectively, was considered.

\begin{table}
 \caption{Thomas-Ehrman (TE) shifts from Fig.~\ref{fig:TE_sep} multiplied by the proton 1$s_{1/2}$ spectroscopic factors (C$^2$S) and added to the energies of states in the mirror nuclei (E$_{\text{mirror}}$) compared to the experimental results (E$_{\text{exp}}$) for \nuc{26,27,28}P. E$_{\text{mirror}}$ energies for all mirror nuclei and E$_{\text{exp}}$ energies for \nuc{28}P were taken from Ref.~\cite{ENSDF}. E$_{\text{exp}}$ energies for \nuc{27}P were taken from Refs.~\cite{SUN1,SUN2}. E$_{\text{mirror+TE}}$ energies are reported relative to the ground state.}
\begin{ruledtabular}
\begin{tabular} {@{}llccccc@{}}
$^\text{A}$Z & J$^{\pi}$ & C$^2$S & TE$\times$C$^2$S & E$_{\text{mirror}}$ & E$_{\text{mirror+TE}}$ & E$_{\text{exp}}$ \\
& & & keV & keV & keV & keV \\
\hline
\noalign{\vskip 1mm}
\nuc{26}P & 3$^+$ & 0.6257 & -254 & 0 & 0 & 0 \\
 & 1$^+$ & 0.4002 & -171 & 82.2 & 165 & 164.4 \\
 & 2$^+$ & 0.4589 & -203 & 232.7 & 284 & 244 \\
 & 2$^+$ & 0.5636 & -259 & 406.7 & 402 &  \\
\hline
\noalign{\vskip 1mm}
\nuc{27}P & 1/2$^+$ & 0.5271 & -169 & 0 & 0 & 0 \\
 & 3/2$^+$ & 0.0956 & -42 & 984.9 & 1112 & 1125 \\
 & 5/2$^+$ & 0.3337 & -177 & 1698.5 & 1691 & 1569 \\
 & 5/2$^+$ & 0.3517 & -201 & 1940.2 & 1908 & 1861 \\
\hline
\noalign{\vskip 1mm}
\nuc{28}P & 3$^+$ & 0.5345 & -129 & 0 & 0 & 0 \\
 & 2$^+$ & 0.3450 & -85 & 30.6 & 75 & 105.6 \\
 & 0$^+$ & 0.4977 & -150 & 972.4 & 951 & 877 \\
 & 3$^+$ & 0.0483 & -15 & 1013.6 & 1127 & 1134 \\
 & 1$^+$ & 0.2895 & -98 & 1372.9 & 1404 & 1313 \\
 & 1$^+$ & 0.2634 & -96 & 1620.3 & 1654 & 1567 \\
 & 2$^+$ & 0.2956 & -107 & 1622.9 & 1645 & 1516 \\
 & 2$^+$ & 0.1755 & -74 & 2138.9 & 2194 & 2104 \\
 & 1$^+$ & 0.0750 & -33 & 2201.4 & 2298 & 2143 \\
 & 4$^+$ & 0.1985 & -87 & 2271.7 & 2313 & 2216 \\
 & 2$^+$ & 0.1798 & -85 & 2486.2 & 2530 & 2406 \\
 & 5$^+$ & 0.2933 & -140 & 2581.8 & 2571 & 2483 \\
\end{tabular}
\end{ruledtabular}
\label{tab:sf} 
\end{table}

The results are summarized in Table~\ref{tab:sf}. Based on this calculation, the 244-keV level in \nuc{26}P observed in this work is more likely to be the first 2$^+$ state than the second. The rms deviation in the experimental energies for mirror states in \nuc{26}P and \nuc{26}Na is 59~keV. After applying the calculated Thomas-Ehrman shift, the deviation between the predicted and experimental level energies in \nuc{26}P is lower at 28~keV. Similarly, for \nuc{27}P and \nuc{27}Mg, the rms deviation for mirror energies is 119~keV and the rms deviation is lowered to 76~keV when accounting for the Thomas-Ehrman effect. However, although including the Thomas-Ehrman shift using this method improves agreement for the first three excited states in \nuc{28}P, the 80-keV rms deviation in mirror energies for \nuc{28}P and \nuc{28}Al is smaller than the 97-keV rms difference between predicted and experimental energies for all excited states in Table~\ref{tab:sf} because the calculated excitation energies for the higher-lying levels in \nuc{28}P are systematically larger than the experimentally observed values by about 110~keV. Here, the mirror energies with no shift are already higher than the experimental energies and including the Thomas-Ehrman effect generally increases the excitation energies further because the Thomas-Ehrman shift is larger in magnitude for the ground state than for nearly all of these excited states, as seen in Table~\ref{tab:sf}.

The same analysis was then repeated for the neighboring \textit{sd}-shell nuclei \nuc{25}Si, \nuc{28}S, and \nuc{29}S. For these nuclei, the $l$=0 proton overlaps with excited states below 4~MeV in \nuc{24}Al, \nuc{27}P, and \nuc{28}P calculatd using USDB were considered. Unlike in \nuc{26,27,28}P above, the summed spectroscopic factor for the one-proton 1$s_{1/2}$ overlaps with the ground state of \nuc{25}Si, which has a proton separation energy of 3414(10)~keV \cite{ENSDF}, is smaller or similar in magnitude to the spectroscopic factors for the excited states. As seen in Table~\ref{tab:sf_n}, the levels in \nuc{25}Na are 267~keV higher than their analog states in \nuc{25}Si on average. After applying the calculated Thomas-Ehrman shift to the levels in the mirror nucleus, the excitation energies are still higher than the experimental values by an average of 183~keV. In this case, the applied Thomas-Ehrman shift is not large enough to fully account for the mirror energy differences. Conversely, for the higher-lying states in \nuc{28}P discussed above, the applied Thomas-Ehrman shift is too large in magnitude.

In USDB calculations for \nuc{28}S and \nuc{29}S, the proton 1$s_{1/2}$ overlaps with states below 4~MeV in \nuc{27}P and \nuc{28}P, respectively, are larger for the ground states than for the excited states, as was true for \nuc{26,27,28}P. The proton separation energies in \nuc{28}S and \nuc{29}S are 2490(160)~keV and 3300(50)~keV \cite{ENSDF}. For \nuc{28}S, applying the Thomas-Ehrman shift calculation yields an excitation energy for the $2^+$ state that is 53~keV below the experimental value while the mirror energy difference is within 34~keV, as shown in Table~\ref{tab:sf_n}. On the other hand, the states in \nuc{29}Al are 126~keV higher in energy than the corresponding levels in \nuc{29}S on average and including the Thomas-Ehrman shift gives predicted energies that are 98~keV higher than the literature values. Therefore, for \nuc{29}S, the magnitude of the Thomas-Ehrman shift would need to be increased to improve agreement further, similar to the case of \nuc{25}Si discussed previously.

An explanation for the remaining discrepancies in the mirror energy shifts may be that the overall rms charge radii of the states gradually increases with excitation energy. For example, the total energy difference between \nuc{28}P and \nuc{28}Al is 11.3 MeV \cite{ENSDF}. Since the Coulomb energy goes as 1/R, a 1\% radius increase would reduce the Coulomb energy by about 110 keV. This increase may be related to the increased occupancy of the 1$s_{1/2}$ orbit as a function of excitation energy. The 1$s_{1/2}$ orbit adds density to the nuclear interior, and as a consequence the nucleus slightly expands in order to keep the interior nuclear saturation density closer to the saturation value of 0.16 nucleons/fm$^3$. For the excited states in Tables~\ref{tab:sf} and \ref{tab:sf_n}, there is moderate correlation (correlation constant of 0.74) between the remaining mirror energy difference after the Thomas-Ehrman shift is applied and the summed 1$s_{1/2}$ occupancy for protons and neutrons relative to the ground state.

\begin{table}
 \caption{Thomas-Ehrman (TE) shifts from Fig.~\ref{fig:TE_sep} multiplied by the proton 1$s_{1/2}$ spectroscopic factors (C$^2$S) and added to the energies of states in the mirror nuclei (E$_{\text{mirror}}$) compared to the experimental results (E$_{\text{exp}}$) for \nuc{25}Si and \nuc{28,29}S. E$_{\text{mirror}}$ energies for \nuc{28}Mg and \nuc{29}Al and E$_{\text{exp}}$ energies for \nuc{28,29}S were taken from Ref.~\cite{ENSDF}. E$_{\text{exp}}$ energies for \nuc{25}Si and E$_{\text{mirror}}$ energies for \nuc{25}Na are the same as those used in Ref.~\cite{LONGFELLOW201897}. E$_{\text{mirror+TE}}$ energies are reported relative to the ground state.}
\begin{ruledtabular}
\begin{tabular} {@{}llccccc@{}}
$^\text{A}$Z & J$^{\pi}$ & C$^2$S & TE$\times$C$^2$S & E$_{\text{mirror}}$ & E$_{\text{mirror+TE}}$ & E$_{\text{exp}}$ \\
& & & keV & keV & keV & keV \\
\hline
\noalign{\vskip 1mm}
\nuc{25}Si & 5/2$^+$ & 0.3088 & -55 & 0 & 0 & 0 \\
 & 3/2$^+$ & 0.2635 & -48 & 89.5 & 97 & 45 \\
 & 1/2$^+$ & 0.4780 & -107 & 1069.3 & 1018 & 870 \\
 & 3/2$^+$ & 0.3613 & -105 & 2202 & 2152 & 1961  \\
 & 9/2$^+$ & 0.2592 & -80 & 2416 & 2391 & 2365  \\
 & 7/2$^+$ & 0.7042 & -230 & 2788 & 2614 & 2380  \\
 & 5/2$^+$ & 0.4634 & -159 & 2914 & 2810 & 2585  \\
 & 7/2$^+$ & 0.2447 & -97 & 3353 & 3312 & 3160  \\
 & 9/2$^+$ & 0.3988 & -159 & 3455 & 3351 & 3035  \\
 & 9/2$^+$ & 0.4324 & -201 & 3995 & 3849 & 3695  \\
 & 1/2$^+$ & 0.4095 & -210 & 4289 & 4135 & 3802  \\
\hline
\noalign{\vskip 1mm}
\nuc{28}S & 0$^+$ & 0.8588 & -187 & 0 & 0 & 0 \\
 & 2$^+$ & 0.6724 & -207 & 1473.5 & 1454 & 1507 \\
\hline
\noalign{\vskip 1mm}
\nuc{29}S & 5/2$^+$ & 0.9049 & -166 & 0 & 0 & 0 \\
 & 1/2$^+$ & 0.7650 & -190 & 1398.1 & 1374 & 1222 \\
 & 7/2$^+$ & 0.6077 & -165 & 1754.3 & 1755 & 1727 \\
 & 5/2$^+$ & 0.6036 & -225 & 3061.8 & 3002 & 2887 \\
\end{tabular}
\end{ruledtabular}
\label{tab:sf_n} 
\end{table}

\section{Summary}
In-beam $\gamma$-ray spectroscopy experiments using the high-efficiency detector array CAESAR and the high-resolution detector array SeGA were performed to measure excited states in the neutron-deficient phosphorus isotopes $^{26,27,28}$P. A level with 244(3)-keV and 80(3)-keV transitions to the ground state and first excited state of the drip-line nucleus \nuc{26}P, respectively, was observed for the first time. Two new measurements of the energy of the $3/2^+$ state, which is the dominant resonance in \nuc{26}Si(p,$\gamma$)\nuc{27}P reaction-rate calculations, were made with the results of 1125(6)~keV and 1119(8)~keV. In \nuc{28}P, $\gamma$ rays from states above the proton separation energy were observed and the lifetimes of the first two excited states were extracted using the $\gamma$-ray lineshape method. The lifetime of the 105.64-keV level was found to be 55(10)~ps and the lifetime of the 877-keV level was found to be 117(30)~ps. The expected Thomas-Ehrman shifts for states in \nuc{26,27,28}P were calculated and applied to known experimental energies in the mirror nuclei. For \nuc{26,27}P, the rms deviation between mirror energies decreased when including the Thomas-Ehrman effect. However, for higher-lying states in \nuc{28}P, the calculated energies are systematically higher than the known energies and no Thomas-Ehrman shift is required. Conversely, for the neighboring nuclei \nuc{25}Si and \nuc{29}S, a larger Thomas-Ehrman shift is needed to better match the experimental data.

\begin{acknowledgments}
This work was supported by the National Science Foundation (NSF) under Grants No.\ PHY-1102511 and PHY-1565546, by the DOE National Nuclear Security Administration through the Nuclear Science and Security Consortium, under Award No.\ DE-NA0003180, and by the Department of Energy, Office of Nuclear Physics, under Grant No.\ DE-FG02-08ER41556. B.A.B. acknowledges support from NSF Grant No.\ PHY-1811855. Discussions with Hironori Iwasaki, Kirby W.~Kemper, and Lewis A.~Riley are acknowledged. 
\end{acknowledgments}

\end{document}